# Detecting local processing unit in drosophila brain by using network theory

Dongmei Shi[1,2], Chitin Shih[3,4], Chungchuan Lo[5], Yenjen Lin[1], and Annshyn Chiang[1,6,7]

1. Brain Research Center, National Tsing Hua University, Hsinchun 30013 Taiwan,
2. Department of Physics, Bohai University, Jinzhou Lianning 121000, P.R.C,
3. Department of Physics, Tunghai University, Taichung 40704 Taiwan,
4. Physics Division, National Center for Theoretical Sciences, Hsinchu 30043, Taiwan,
5. Institute of Systems Neuroscience, National Tsing Hua University, Hsinchu 30013, Taiwan,
6. Institute of Biotechnology, National Tsing Hua University, Hsinchu 30013, Taiwan
7. Genomics Research Center, Academia Sinica, Taipei 11529, Taiwan

Community detection method in network theory was applied to the neuron network constructed from the image overlapping between neuron pairs to detect the Local Processing Unit (LPU) automatically in Drosophila brain. 26 communities consistent with the known LPUs, and 13 subdivisions were found. Besides, 45 tracts were detected and could be discriminated from the LPUs by analyzing the distribution of participation coefficient P. Furthermore, layer structures in fan-shaped body (FB) were observed which coincided with the images shot by the optical devices, and a total of 13 communities were proven closely related to FB. The method proposed in this work was proven effective to identify the LPU structure in Drosophila brain irrespectively of any subjective aspect, and could be applied to the relevant areas extensively.

**Instruction**

Some neuron circuits in Drosophila are believed worth studying in understanding the fundamental features of neurons by most neuroscientists [1-4]. Genetic screens in Drosophila have identified many genes involved in neural development and function [5-9]. One reason is the power of the Drosophila genetic toolbox [10-12], another basic reason is that fruit flies have about 1000-fold fewer neurons, while exhibit extensive cognitive abilities.

Local processing unit (LPU) [13-14] was proposed in constructing a mesoscopic map of the brain circuit in Drosophila. A LPU is defined as a brain region consisting of its own local interneurons (LN) whose nerve fibers are completely restricted to this region. Further, each LPU is contacted by at least one nerve tract. LPU is a functional unit dealing with the signals, which plays a critical part in information transmission in the brain. Connectomics-Based analysis of information flow in the drosophila brain based on the LPUs has been studied recently, and it was shown that the network showed hierarchical structure, small world, and rich-club organization [14]. Normally, LPU and its boundary are determined by observation based on the optical devices, which exists some subjective factors.

In this work, complex network theory [15-18] was introduced to detect the LPU structure in Drosophila brain based on LPU's definition. It was proven that LNs and neuron tracts could be both identified successfully, and so that LPU would be determined by LNs and the corresponding neuron tracts, irrespectively of any subjective aspect.

## Methods

Community structure (Figure 1) [19]: A network is realized to have the community structure if the nodes of network can be easily grouped into sets of nodes such that each set of nodes is densely connected internally.

Modularity Q: Modularity is one measure of the structure of networks or graphs. It was designed to measure the strength of division of a network into modules (also called groups, or communities).

$$Q = \frac{1}{m} \sum_{i,j} \left[ A_{ij} - \frac{s_i^2}{2m} \right] \delta(c_i - c_j),$$

where $m = \sum_{ij} A_{ij}$, and $c_i$ is the module which the node i belongs to. $s_i$ is the strength of node i, and $A_{ij}$ is the element in adjacency matrix. $c_i$ is determined by maximizing the modularity Q, and larger Q indicates more distinct community structure in the networks.

Participation coefficient P:

$$P_i = 1 - \sum_{j=1}^{N_c} \left( \frac{s_{ij}}{s_i} \right)^2, (0 \leq P_i \leq 1),$$

where $s_{ij}$ is the strength of node i connected to the nodes in module $c_j$, and $s_i$ is the total strength of node i in the network. $P_i$ is close to 0 if most of the edges node i linked to are in its own module $c_i$. On the other hand, $P_i$ is close to 1 if the links are uniformly distributed among all the modules of the network.

## Simulations

Neuron network construction: Neuron network was based on the 23380 female fruit fly data. The nodes in the network indicated the neurons of brain, and the links represented the interactions among the neurons. The corresponding adjacency matrix was constructed as follows: the matrix dimension was the total number of neurons, and the matrix element $A_{ij}$ was the overlapping strength in space of neuron i and neuron j. The neuron network was an undirected-weighted complex network.

Relations between LPU in the brain neuron system and community structure in a network (Figure 2): Neurons in the fruit fly brain could be categorized into two functional distinct populations:

local interneurons (LNs) whose processes were restricted within a single brain region, and projection neurons (PNs) whose dendrites and axons connected two or more brain regions. Local Processing Unit (LPU) was defined as a brain region consisting of its own LNs population whose nerve fibers were completely restricted to the region, and meanwhile, each LPU was connected by at least one neural tract. So LPU boundary could be determined by LNs and the neuron tract fibers restrained to this region. On other hand, according to the definition of community structure in a network, two types of nodes could be classified: nodes whose links were all in a single community (P=0), and nodes whose links were distributed into several communities (P>0). Accordingly, a correspondence could be built up between LPU in the brain and community structure in a network: the LNs corresponded to the nodes with P=0, and PNs corresponded to the nodes with P>0. Figure 2 showed the representation of this relation.

Community division: Based on the analysis above, detecting LPU structure was tantamount to find or form the communities in which participation coefficient P for all the nodes met P=0. Firstly, modularity calculation was used to divide the neuron networks into communities. In order to detect the basic LPU structure in which no smaller LPUs existed, modularity calculation was performed three times sequentially based on the original neuron network. Precisely (Fig. 3), the original network was divided into 8 communities by calculating Q (Cal. Q), and these 8 communities were called level 1 communities (L1-Comm). Then, L1-Comm would be further divided into smaller communities which were called level 2 communities (L2-Comm), and further, L2-Comm would be done by the same analogy into level 3 communities (L3-Comm). Taking the 5th community on level 1 for example, it was seen in Fig. 3 that the 5th community on L1-Comm was divided into 3 small communities on level 2, 5-1, 5-2, 5-3. And the second community 5-2 on level 2 could be further divided into 3 smaller communities on level 3, 5-2-1, 5-2-2, 5-2-3. And so on, for each of the 8 communities on level 1.

Community detection method: Representation of community detection method was shown in Fig. 4, one can see that three steps were needed to detect the LPU structure. On the first step, the original brain network was divided into L3-Comm communities by Cal. Q, among which LPU communities were included. On step 2, neuron tract communities would be discriminated. It had been proven according to the statistical results in this study that a community was the tract community if $P_{min} \geq 0.3$, or if there existed a distinct segmentation point $p^* \sim 0.5$ or P>0.5 (f(P)~0) in the distribution of P, then the neurons corresponding to the last segmentation belonged to a neuron tract. $P_{min}$ is the minimum of participation coefficient in this community, and f(P) is the distribution function of P. See an example of tract distribution in Fig. 5.

LPU communities were left by manipulations in step 2, but there still existed PNs in these communities. In order to extract LNs from LPU communities, LPU community would be optimized in step 3. Based on correspondence between LPU and community structure (Fig. 2), our solution was to remove the nodes with large P piece by piece from the LPU community by modularity calculating successively, until $\bar{P} \leq 0.1$ of this community. The community could be realized isolated when $\overline{P} \leq 0.1$ in our study.

Before our further study, there were two points needed to be discussed. According to the theory in community detection method, the nodes with participation coefficient P=0 corresponded to the LNs whose fibers were all restricted to one region. However, considering the existence of warping area differences, when the neurons were put into the standard brain [20-21], the critical value of P could not reach 0 accurately. Actually P should be a function of warping area, and was different in different brain regions.

On other hand, considering that some LPUs might be destroyed by the successively divisions in step 1, operations on step 3 would be applied directly to the communities in level 1 to compensate the destroyed LPUs.

**Results**

Community detection method was applied to the Drosophila brain neuron system by means of the above manipulations. Figure 6 showed the LNs of different brain regions in the standard brain from forward, middle and backward levels of the brain, and it was proven that these LNs belonged to 26 known LPUs: DLP (dlp), MB (mb), AL (al), VLP (vlp), MED (med), LOB (lob), nod, FB, EB, LH (lh), LOP (lop), PB, CMP (cmp). Two subdivisions in VLP (vlp), three subdivisions in FB, and three subdivisions in MB (mb) were also illustrated. Generally, 49 LPUs and hubs had been found, and the lost LPUs in this work were mainly due to the fewer LNs supplied in these LPUs.

45 neuron tracts were discriminated by the community detection method, and were shown in Fig. 7.  9 types of neuron tracts were proven connected AL, which were sighed 1-9 in Fig. 8.

Subdivisions in FB and MB were shown in Fig. 8 (1) and Fig. 8 (2). Three subdivisions in FB were detected which had been not observed in experiment, but were proven to be consistent with the results only based on FB data by applying the community detection method. Three subdivisions ($\alpha$, $\beta$, $\gamma$) in MB (mb) shown in Fig. 8 (2) coincided with the observations in experiment by optical devices.

Four layers in FB were detected, and illustrated in Fig. 9. Only neurons in Fig. 9 (3) were the LNs, and neurons in Fig. 9 (1), (2), (4) were all PNs according to the community detection method.

Figure 10 shows the communities related to FB, including three LPU communities and other 10 tract communities. Apparently FB played an important part in information transmission in the brain.

**Summary**

We applied the community detection method to detect the LPUs in Drosophila brain system, and had proven that 26 known LPUs were detected, and subdivisions were also found. Precisely, three subdivisions in MB (mb), and two subdivisions in VLP (vlp) were consistent with the observations in experiments, and unknown three subdivisions in FB would give us important information to understand FB's internal structure. It was further proven that the same subdivisions in FB were also found if the community method was only applied to FB neuron system. Besides, layer structures in FB were detected, and one layer was a subdivision in FB, and other three layers were tract communities. A total of 13 communities were closely related to FB, including subdivisions of FB, neuron tract communities inside FB, and neuron tract communities on FB's boundary. Apparently, FB played an important part in information transmissions in Drosophila brain. Moreover, it had been proven that neuron tracts could be discriminated by analyzing the distributions of participation coefficient P.

Community detection method was proven effective to detect LPU in Drosophila brain system in this study, and its boundary was determined by LNs and the neuron tracts totally, irrespectively of any subjective factor. This method would be more perfect if the participation coefficient P as a function of warping area was known, and would be applied to the relative areas extensively.

**Appendix** [Detecting subdivisions in FB and subdivisions in neuron tract communities]
Manipulations in step 3 of community detection method (Fig. 4) were directly applied to FB neuron system in which 544 neurons in FB had been obtained by experiment. Precisely, nodes with large P would be removed piece by piece from the FB neuron network successively until $\overline{P} \leq 0.1$ of this community. It had been proven that three subdivisions were found in FB coinciding with the results in the above simulations. Moreover, this method was also applied to neuron tract communities discriminated in step 2 in Fig. 4 in order to find the basic neuron tracts in these communities.

**Acknowledgments**
This work was supported by the MOE 5-Year-50-Billion Project and NSC Nano National Project in Taiwan, Specialized Foundation for Theoretical Physics of China (Grant No. 11247239), National Natural Science Foundation of China (Grants No. 11305017).

Figure 10. 13 communities closely related to FB.

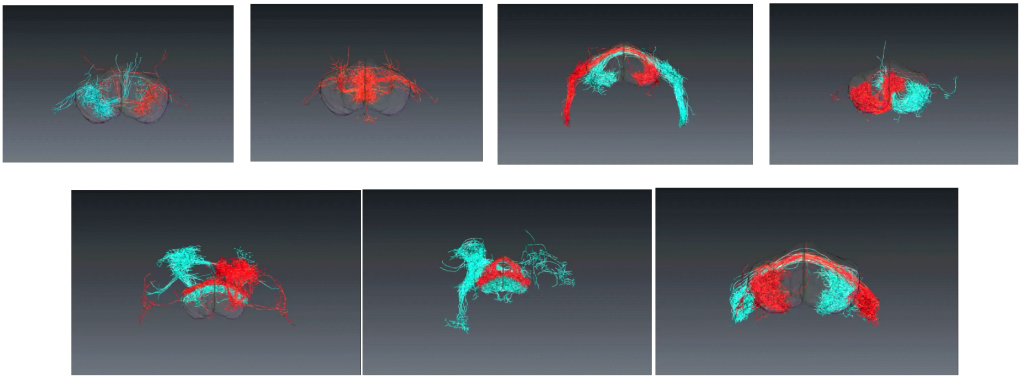

Figure 9. Layer structures in FB.

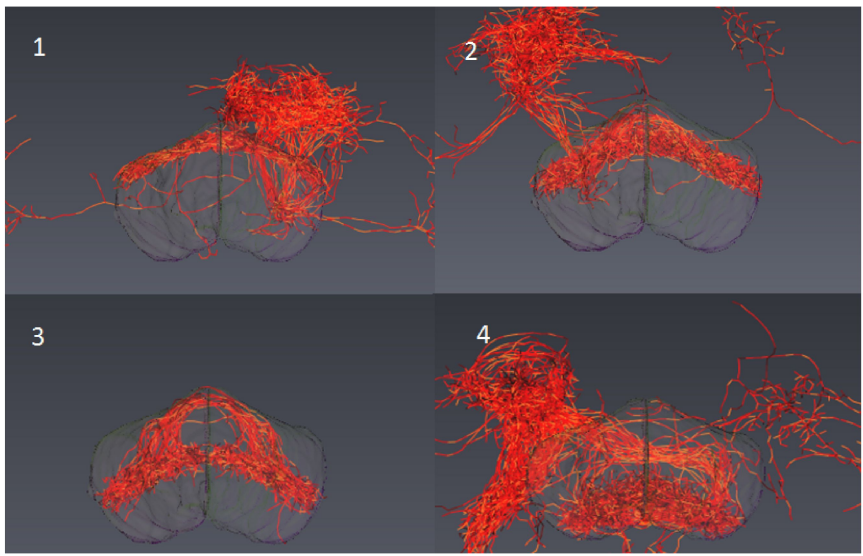

Figure 8. (1) Subdivisions in FB; (2) Subdivisions in MB (mb).

(1)

(2)

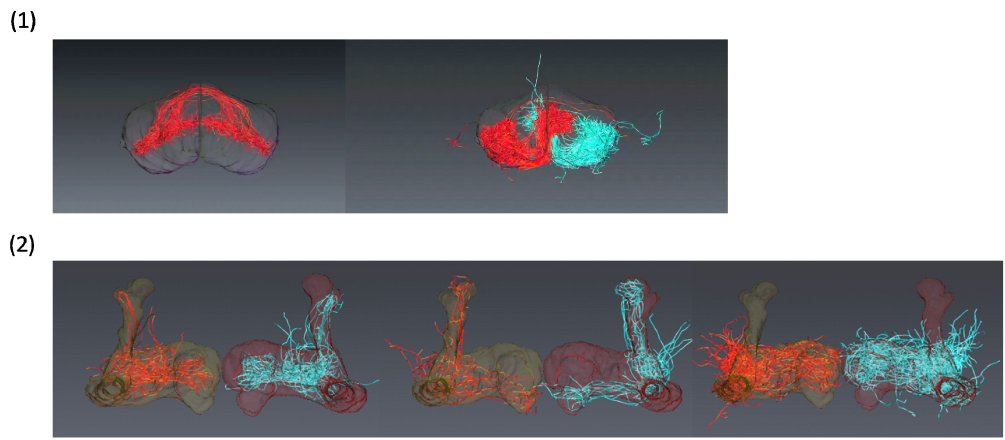

Figure7. 45 neuron tracts in the standard brain, and neuron tracts signed 1-9 were the tracts connecting the AL (al).

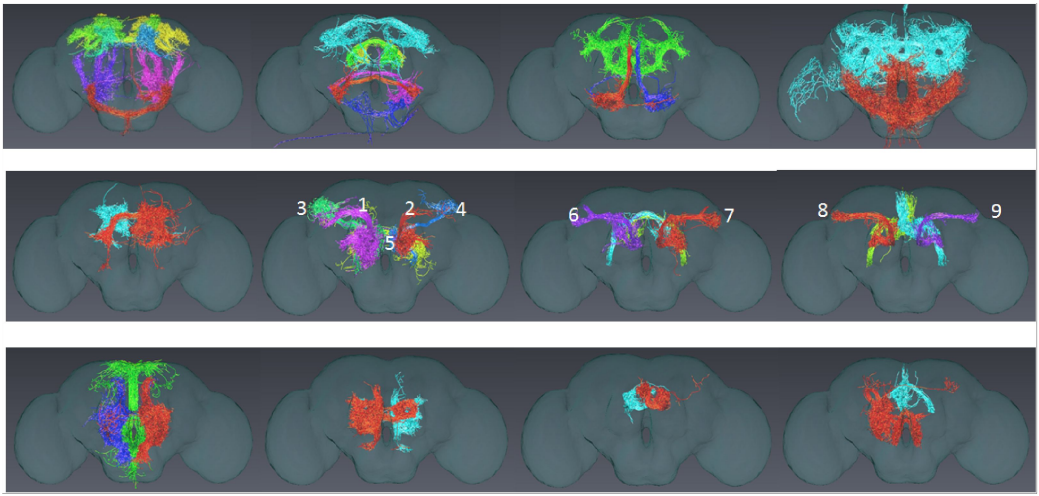

Figure6. LNs in different regions in the standard brain from forward, middle and backward level, respectively.

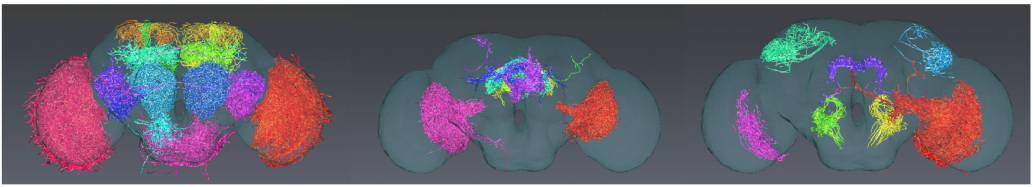

Figure5. Tract community discrimination.

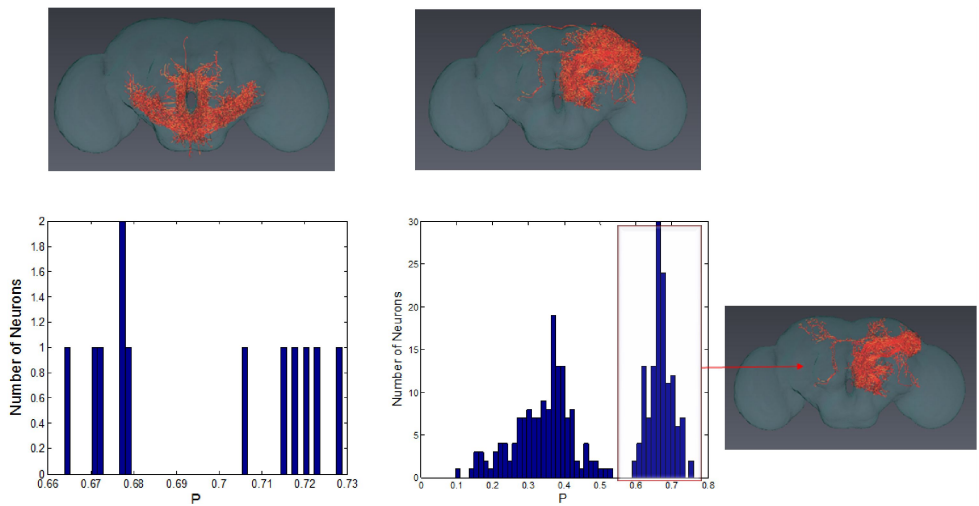

Figure 4. Graphic representation of community detection method. Step 1, community division; step 2, neuron tract discrimination; step 3, LPU community optimization.

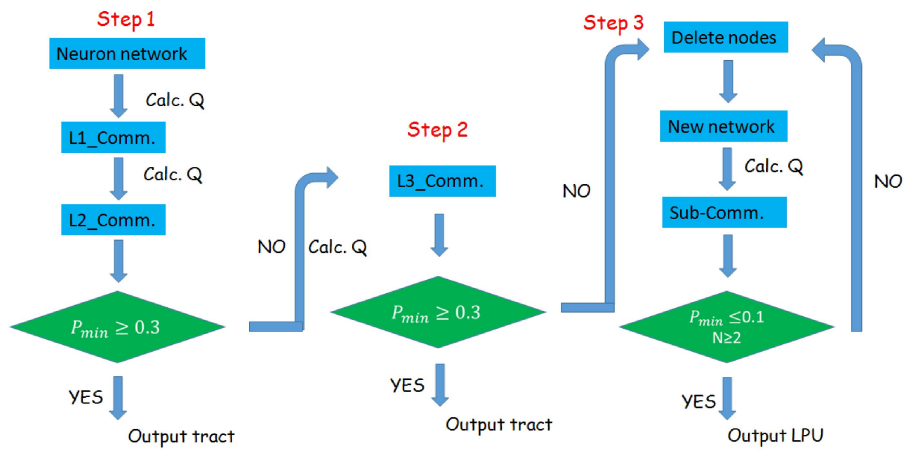

Figure3. Community division by calculating modularity ($Calc.\ Q$). The original brain neuron system was divided into 8 communities called level 1 community ($L1-Comm$), and these 8 communities could be further divided into sub-communities called level 2 community ($L2-Comm$). Based on $L2-Comm$, they would be further divided to communities called level 3 community ($L3-Comm$). Taking the 5th community on level 1 for example, it was divided into 3 smaller communities 5-1,5-2,5-3 on level 2, and community 5-2 was further divided into 3 sub-communities on level 3.

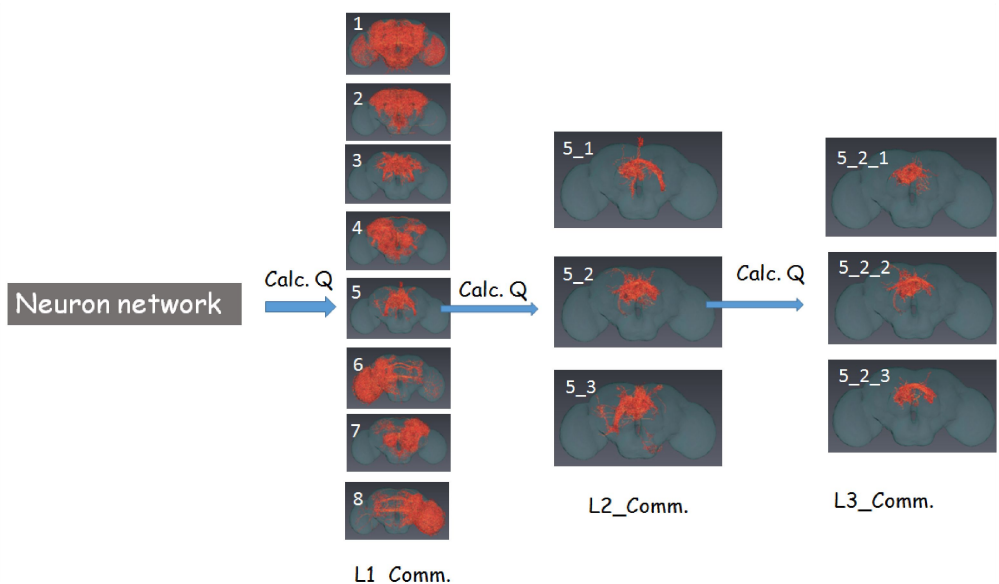

Figure 2. Relations between LPU and community. (1) Black-solid curves represented the local interneurons (LNs), and other color-solid curves indicated the projection neurons (PNs). Green, blue, and red-solid curves represented neuron tracts; (2) Nodes in the same shadow formed a community, and two types of nodes were included, type of node A whose links were all in a single community ($P_{A}=0$), and type of B whose links were distributed into one more communities ($P_{B}>0$).

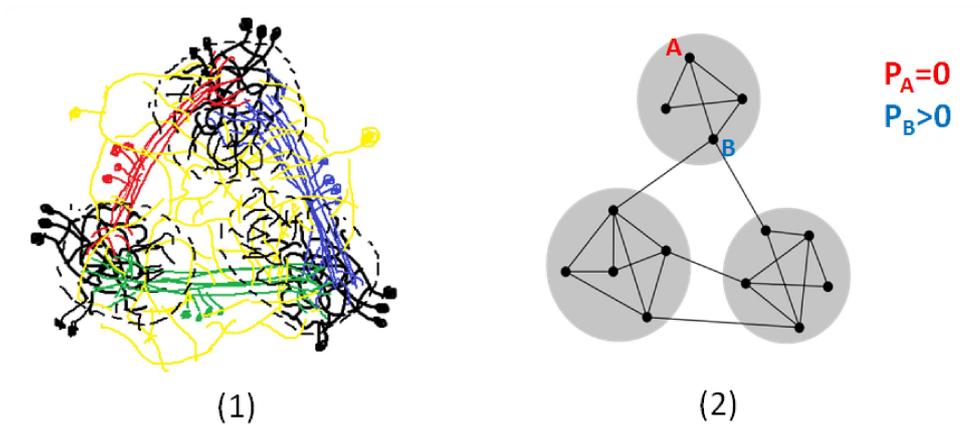

(1)　　　　　　　　　　(2)

Figure 1. Representation of community structure in a graph. Nodes in the same shadow form a community.

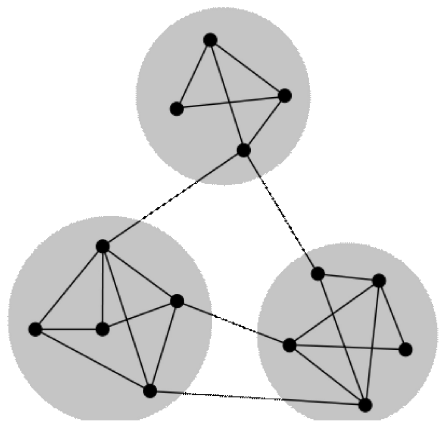